\documentclass[%
 reprint,
 amsmath,amssymb,
pra,
]{revtex4-1}

\usepackage{graphicx}% Include figure files
\usepackage{dcolumn}% Align table columns on decimal point
\usepackage{bm}% bold math
\usepackage{natbib}
\usepackage{siunitx}
\usepackage{amsmath}
\usepackage[utf8]{inputenc}
\usepackage{comment}
\usepackage{filecontents}

\makeatletter
    \let\@internalcite\cite
    \def\cite{\def\citeauthoryear##1##2{##1, ##2}\@internalcite}
    \def\shortcite{\def\citeauthoryear##1{##2}\@internalcite}
    \def\@biblabel#1{\def\citeauthoryear##1##2{##1, ##2}[#1]\hfill}
\makeatother

\begin{document}

\preprint{APS}

\title{Topological edge and corner states \\ in a two-dimensional photonic Su-Schrieffer-Heeger lattice}% Force line breaks with \\

\author{Minkyung Kim}
\author{Junsuk Rho}%
 \email{jsrho@postech.ac.kr}
 \altaffiliation[Also at ]{Department of Chemical Engineering, Pohang University of Science and Technology (POSTECH), Pohang 37673, Republic of Korea}
\affiliation{Department of Mechanical Engineering, Pohang University of Science and Technology (POSTECH), Pohang 37673, Republic of Korea}


\begin{abstract}
Implementation of topology on photonics has opened new functionalities of photonic systems such as topologically protected boundary modes. We present polarization-dependent topological properties in 2D Su-Schrieffer-Heeger lattice by using a metallic nanoparticle array and considering the polarization degree of freedom. We demonstrate that when eigenmodes are polarized parallel to the plane of the 2D lattice, it supports isolated longitudinal edge modes and transverse modes that are hidden from the projected bulk states. Also, the in-plane polarized modes support a second-order topological phase under an open boundary condition by breaking the four-fold rotational symmetry. This work will offer polarization-based multifunctionality in compact photonic systems that have topological features.
\begin{description}
\item[PACS numbers]
42.25.Ja, 42.70.Qs, 42.79.Sz
\end{description}
\end{abstract}

\pacs{Valid PACS appear here}% PACS, the Physics and Astronomy
\maketitle

%\tableofcontents
\renewcommand{\vec}[1]{\mathbf{#1}}
\section{Introduction}
Introduction of the concept of topology to band theory has not only enriched our understanding of phases of matter, but also spawned a new field called topological band theory, which has explained many anomalous behaviors such as the quantum Hall effect \cite{klitzing1980new} and the quantum spin Hall effect \cite{kane2005quantum}. Such exotic phenomena have been extended and reproduced in classical systems by means of photonic/phononic crystals \cite{PhysRevLett.114.223901, he2016acoustic, PhysRevLett.114.114301, PhysRevLett.100.013905, yang2019realization}, metamaterials \cite{khanikaev2013photonic, PhysRevLett.114.037402, yang2017direct, doi:10.1002/adom.201900900, PhysRevB.99.235423} and circuitry \cite{imhof2018topolectrical, PhysRevX.5.021031}. Among the many models supporting the topological phases, the Su-Schrieffer-Heeger (SSH) model, a dimerized chain, is known as a long-standing and the simplest model \cite{PhysRevLett.42.1698}. The 1D SSH model and its generalization to 2D have been extensively investigated in various areas including electronic systems, photonics and acoustics for ample physics despite the simple structures such as higher-order topological phase \cite{PhysRevLett.122.233902, PhysRevB.98.205147, PhysRevLett.122.233903} and fractional charge \cite{RevModPhys.60.781}.

Here, we present the polarization-dependent topological properties of a 2D photonic SSH model consisting of metallic nanoparticles (NPs). In general, for 2D models, eigenmodes that are polarized perpendicular to the plane of the 2D lattice are considered. However, polarization in photonics can provide an additional degree of freedom to independently control topological features such as transport phenomena without requiring the structural modification. We demonstrate that the 2D SSH model possesses topological phase characterized by a 2D Zak phase also for polarization parallel to the plane of the lattice, thereby supporting longitudinal edge modes. Furthermore, by exploiting the polarization dependence to break $C_4$ symmetry of the 2D SSH model, a second-order topological phase, which is a 0D corner mode in this 2D system, is observed under an open boundary condition, whereas previous demonstrations of second-order topological features require an interface between expanded and shrunken 2D SSH models \cite{PhysRevLett.122.233902, PhysRevB.98.205147, PhysRevLett.122.233903}. This relaxed condition of the second-order topological phases and polarization-based control of topological states will be advantageous in realizing compact photonic platform for robust manipulation of light.

\section{Bulk states}
\begin{figure}[h!] \centering
\includegraphics [width=0.45\textwidth]{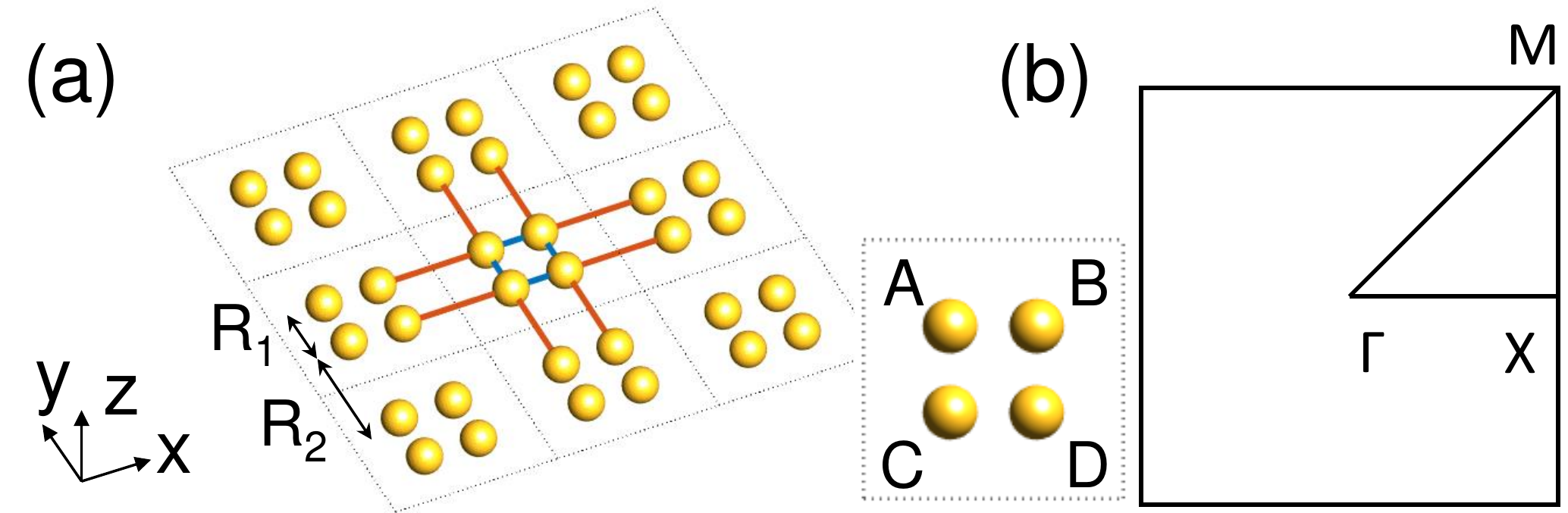}
\caption{(a) Schematic of a 2D photonic SSH model composed of NPs. Four sites of a unit cell are labelled as $A, B, C$ and $D$. Blue and red lines indicate intracellular and intercellular coupling respectively. $R_1$ and $R_2$ represent  the intracellular and intercellular distance. (b) The first Brillouin zone with high symmetry lines.
}
\label{schematic}
\end{figure}

Four NPs compose a unit cell of a 2D photonic SSH model (Fig. \ref{schematic}(a)). The intracellular and intercellular distance will be represented as $R_1$ and $R_2$ respectively. The first Brillouin zone (BZ) of the 2D SSH model is a square as shown in Fig. \ref{schematic}(b).

To build an eigenvalue problem, we use a coupled dipole approximation method \cite{PhysRevLett.82.2776, yurkin2007discrete, honari2019topological}. The NPs at $\vec{R}$ can be described by a point dipole $\vec{p(\vec{R})}$ if the radius of the NPs is much smaller than the center-to-center distance \cite{PhysRevB.69.125418}. Under linear and nonmagnetic assumption, the dipole moment $\vec{p}_n$ of a particle positioned at $\vec{R}_n$ can be calculated by multiplying polarizability by a summation of the total electric fields induced by other dipoles. Then, the self-consistent dipole of an array of particles can be expressed as \cite{RevModPhys.79.1267}.
\begin{equation}
    \label{dipole_sum}
    \vec{p}_n = \alpha_E \sum_{m \neq n}G^{0}(\vec{R}_n-\vec{R}_m)\vec{p}_m
\end{equation}
where $\alpha_E$ is a electric polarizability tensor of the NPs given as $\alpha_E = 4\pi\varepsilon_0 r^3 (\varepsilon - \varepsilon_g)/(\varepsilon + 2\varepsilon_g)$. Here, $\varepsilon_g$ is the background permittivity, and the permittivity $\varepsilon$ of the NPs follows the Drude model $\varepsilon(\omega) = \varepsilon_\infty - \omega^2_p/(\omega^2+i\gamma\omega)$ for plasma frequency $\omega_p$ and damping frequency $\gamma$; $\varepsilon_\infty$ is the permittivity when frequency goes to infinity. Under the quasi-static approximation, where the wavelength of interest is much larger than the lattice constant, the dipole-dipole interaction tensor is \cite{RevModPhys.79.1267}
\begin{equation}
    \label{G}
    G^0(\vec{R})\vec{p} =  -\frac{1}{R^3}\vec{p} + \frac{3}{R^5}(\vec{R}\cdot \vec{p})\vec{R}
\end{equation}

Since the strength of the dipole-dipole interaction attenuate rapidly as the separation $R$ increases ($\propto 1/R^{3}$), we consider only nearest-neighbor coupling. Then for any $n$ and $m$, the difference $\vec{R}_n-\vec{R}_m$ in Eq. \ref{dipole_sum} has only an $x$-component or a $y$-component. We can further simplify $G^0(\vec{R})$ to $G^0(\vec{R}) = 2/R^3$ $(-1/R^3)$ if $\vec{R}$ is parallel (perpendicular) to the dipole moment direction. Therefore, the electric field induced by a dipole has a polarization parallel to the dipole moment. Because $x$-, $y$- and $z$-polarized modes are fully independent, the eigenvalue problem can be decoupled to three equations of 4 by 4 matrices instead of one equation of a 12 by 12 matrix. This is a direct analogy to the tight-binding model, for which the bulk Hamiltonian can be written as
\begin{align}
    \label{H}
    H_0 & =  t_x(B^{+}_{m,n}A_{m,n} + D^{+}_{m,n}C_{m,n}) \notag \\
    & + t_y(A^{+}_{m,n}C_{m,n} + B^{+}_{m,n}D_{m,n}) \notag \\
    & + \bar{t}_x(A^{+}_{m+1,n}B_{m,n} + C^{+}_{m+1,n}D_{m,n}) \notag \\
    & + \bar{t}_y(C^{+}_{m,n+1}A_{m,n} + D^{+}_{m,n+1}B_{m,n}) + \text{h.c.}
\end{align}
where h.c. denotes Hermitian conjugate; $A^{+}_{m,n} (A_{m,n})$ is a creation (annihilation) operator on a site A at $(m,n)$-th unit cell; $t_i (\bar{t}_i)$ is an intracellular (intercellular) coupling strength in the $i = {x, y}$ direction. For x-polarized mode, the coupling strengths are $t_x = 2/R^3_1$, $t_y = -1/R^3_1$, $\bar{t}_x = 2/R^3_2$ and $\bar{t}_y = -1/R^3_2$. Coupling strengths for y-polarized mode can be obtained straightforwardly in a similar way. For z-polarized mode, $t_x = t_y = -1/R^3_1$ and  $\bar{t}_x = \bar{t}_y = -1/R^3_2$. Hereafter, we refer to $x$ and $y$ polarization ($z$ polarization) as in-plane (out-of-plane) polarization. By applying the Bloch theorem, we can construct a 4 by 4 matrix of a bulk Hamiltonian for each polarization as
\begin{align}
    \label{Hx}
    H & = 
    \begin{pmatrix}
    0 & H_{12} & H_{13} & 0 \\
    H^{*}_{12} & 0 & 0 & H_{24} \\
    H^{*}_{13} & 0 & 0 & H_{34} \\
    0 & H^{*}_{24} & H^{*}_{34} & 0
    \end{pmatrix}, \notag \\
    H_{12} & = H_{34} = \bar{t}_x e^{ik_x R_2} + t_x e^{-ik_x R_1} \notag \\
    H_{13} & = H_{24} = t_y e^{ik_y R_1} + \bar{t}_y e^{-ik_y R_2} %\notag \\
    %H_{24} & = t_y e^{ik_y R_1} + \bar{t}_y e^{-ik_y R_2} %\notag \\
    %H_{34} & = \bar{t}_x e^{ik_x R_2} + t_x e^{-ik_x R_1}
\end{align}

\begin{figure*} \centering
\includegraphics [width=0.75\textwidth]{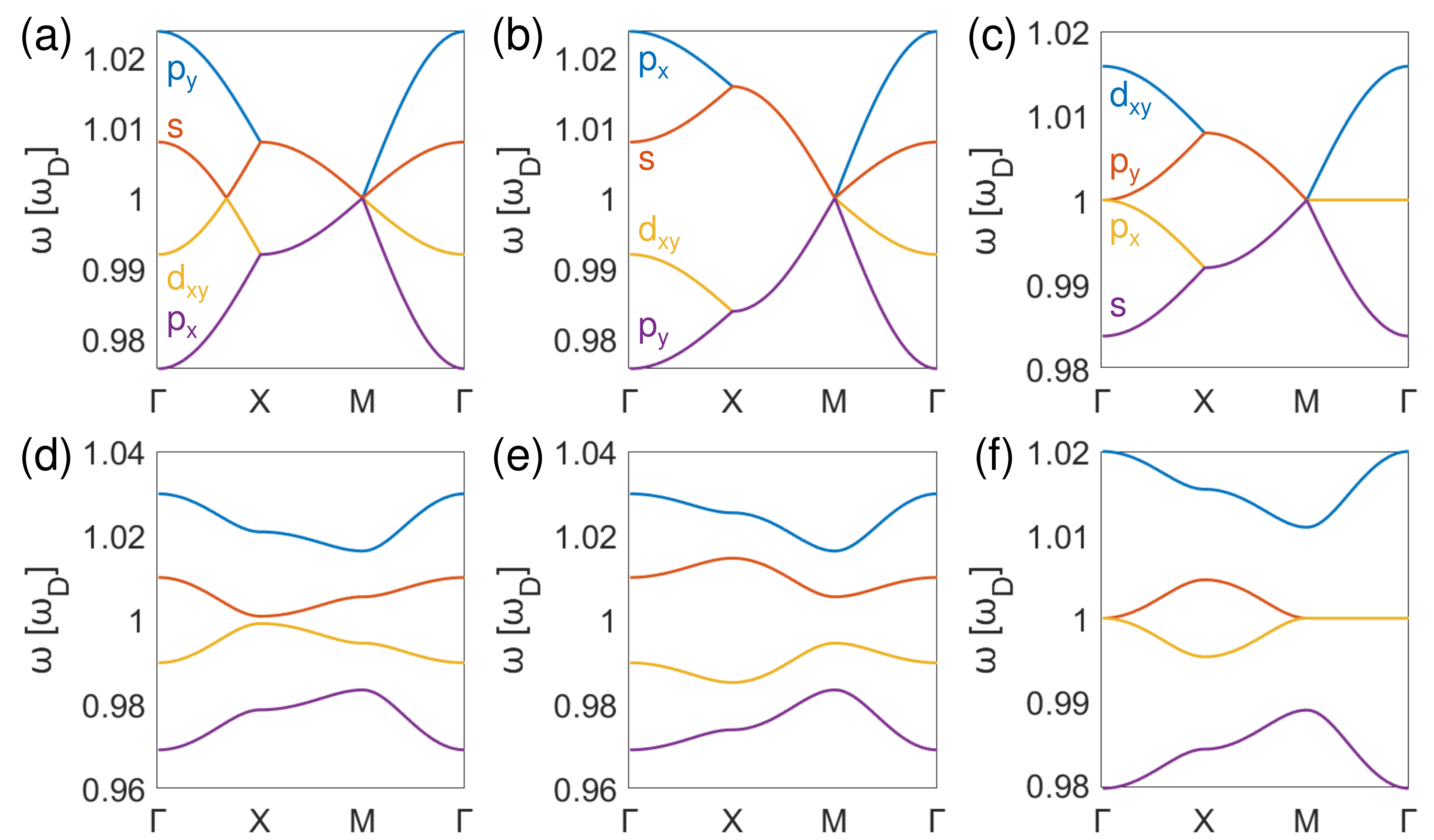}
\caption{Bulk states along high symmetry lines when (a)-(c) $R_1 = R_2$ and (d)-(f) $R_1 \neq R_2$. Left, center and right column correspond to $x$-,$y$- and $z$-polarized modes, respectively.
}
\label{bulk}
\end{figure*}

\begin{figure*} \centering
\includegraphics [width=0.8\textwidth]{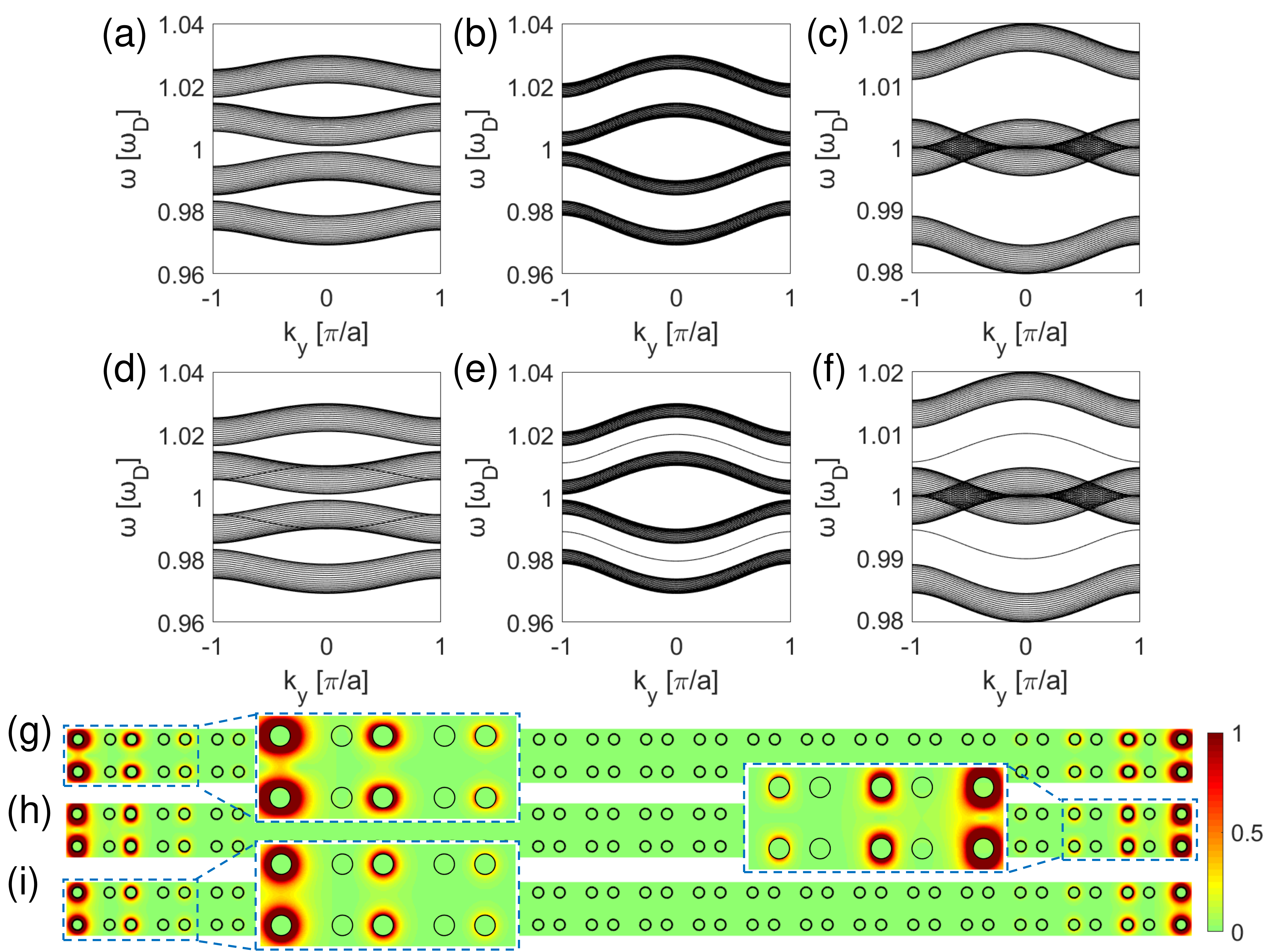}
\caption{(a)-(c) Projected bulk states and edge states of 20 unit cells aligned along $x$-axis. Left, center and right column correspond to $x$-,$y$- and $z$-polarized modes, respectively. (a)-(c) $R_1 < R_2$ and (d)-(f) $R_1 > R_2$. (g)-(i) Normalized electric field distributions of $x$-,$y$- and $z$-polarized edge modes respectively.
}
\label{edge}
\end{figure*}

Bulk dispersion of $x$, $y$ and $z$ polarization along the high symmetry line (Fig. \ref{schematic}(b)) are calculated by solving the eigenvalue problem $\alpha^{-1}_{E}\vec{p}_i = H\vec{p}_i$ where $\vec{p}_i$ is a 1 by 4 vector consisting of the $i$-th component of the dipole moments of four sites in a unit cell for $i = x,y,z$ (Fig. \ref{bulk}). Geometrical parameters are set as: lattice constant $a = 100$ nm, radius of NP $r = a/10, R_1 = 0.6a$ for $R_1 > R_2$ and $R_1 = 0.4a$ for $R_1 < R_2$, and $R_2 = a-R_1$ We use the Drude parameter of gold taken from \cite{blaber2009search} for NPs: $\varepsilon_\infty = 1, \omega_p = 2.07 \times 10^{15} \text{s}^{-1}$ and $\gamma = 4.45 \times 10^{12} \text{s}^{-1}$. The eigenfrequency is normalized by the Dirac frequency of the NP: $\omega_D = \omega_p/\sqrt{2\varepsilon_g+\varepsilon_\infty}$. For given parameters, $\omega_D = 1.20 \times 10^{15} \text{s}^{-1}$ which corresponds to wavelength of $1.58 \mu m$.

First we focus on cases where NPs are equally spaced ($R_1 = R_2$, Fig. \ref{bulk}(a)-\ref{bulk}(c)). Geometrically, the 2D SSH model possesses $C_4$ symmetry in $x$-$y$ plane. If we disregard the symmetry breaking due to the in-plane polarization, bulk states also have $C_4$ symmetry, and in such cases, spatial field distributions of bulk states at $\Gamma$ are associated with $s$, $p_x$, $p_y$ and $d_{xy}$ bands from lower to upper band in order. However, for $x$- and $y$-polarized modes, in-plane polarization breaks the symmetry, and this breaking leads to an inversion of such states. For $x$ polarization, the lowest and the highest bands have $p_x$ and $p_y$ states respectively, whereas the other bands have $s$ and $d_{xy}$ at $\Gamma$ (Fig. \ref{bulk}(a)). Such states become hybridized as the wave vector moves from $\Gamma$ to $X$, and thereby form two doubly degenerate states along $X$-$M$. The degeneracies along $X$-$M$ when $R_1 = R_2$ are a consequence of BZ folding that can occur because the unit cell was set to be reducible. However, along $M$-$\Gamma$, the degeneracies are lifted as a result of the broken $C_4$ symmetry. Bulk states of the $y$ polarization can be analyzed similarly with $p_x$ and $p_y$ switched (Fig. \ref{bulk}(b)). In contrast, bulk states for $z$ polarization show the same features as the conventional 2D SSH model, such as $s$, $p_x$, $p_y$ and $d_{xy}$ from lowest to highest bands at $\Gamma$, and degeneracy of $p_x$ and $p_y$ bands along $\Gamma$-$M$ line (Fig. \ref{bulk}(c)). When $R_1 \neq R_2$, all bands are gapped except the $z$-polarized $p_x$ and $p_y$ bands, in which degeneracy is protected by $C_4$ symmetry. We do not specify whether $R_1$ or $R_2$ is largest, because sub-lattice symmetry guarantees the same bulk states when $R_1$ and $R_2$ are interchanged. However, parities of the spatial distribution of dipole moments change when $R_1$ and $R_2$ are switched, indicating a topological phase transition.

\section{Edge and corner states}
\begin{figure} \centering
\includegraphics [width=0.45\textwidth]{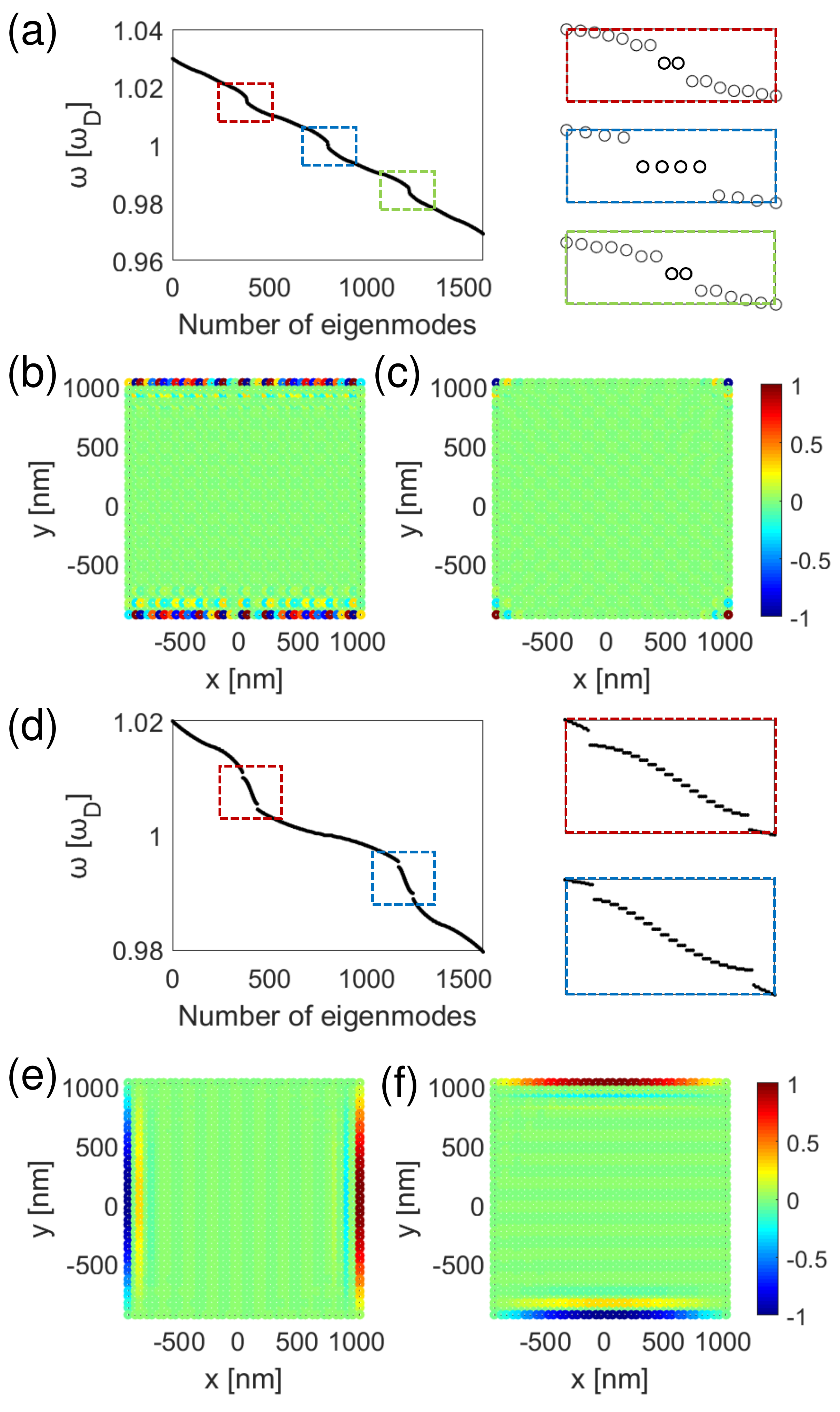}
\caption{Eigenmodes of an array of 20 by 20 unit cells. (a) Eigenfrequency of $x$ polarization and magnified images of edge (red and green boxes) and corner (blue box) states. Normalized dipole distribution of (b) an edge mode and (c) a corner mode. (d) Eigenfrequency of $z$ polarization and magnified views of edge states. (e) and (f) Normalized dipole distribution of two distinct edge modes.
}
\label{NbyN}
\end{figure}

To simulate the edge states, we solve the eigenvalue problem of an array of 20 unit cells aligned along the $x$-axis with periodic boundary assumption along the $y$-axis. In this case, the boundary is an interface between the NP arrays and air, i.e., parallel to the $y$-axis. When $R_1 < R_2$, the topological phase is trivial, and no edge states are found (Fig. \ref{edge}(a)-\ref{edge}(c)). In contrast, when $R_1 > R_2$, edge states appear in all three polarization (Fig. \ref{edge}(d)-\ref{edge}(f)), among which $x$-polarized edge states overlap with the projected bulk states. Inversion symmetry makes the edge states doubly degenerate, with phases symmetric and anti-symmetric along the aligned direction. Normalized electric field amplitudes of edge modes at $k_y$ = 0.5 are localized at boundaries (Fig. \ref{edge}(g)-\ref{edge}(i)). Despite the existence of $x$-polarized edge states, it is impossible to selectively excite edge modes while keeping the bulk insulating because of the overlap of the edge and bulk modes (Fig. \ref{edge}(d)). In short, the in-plane polarized 2D SSH model supports isolated longitudinal edge modes and hidden transverse edge modes. The conventional 2D SSH model, in which out-of-plane polarization is assumed, can also support edge modes along either $x$- or $y$-axis by implementing anisotropy \cite{PhysRevB.98.205147}. However, by considering the in-plane polarization, edge modes with specified propagating direction can be selectively excited depending on the polarization without needing to change the geometrical structures.

Interestingly, in-plane polarization in the 2D SSH model also supports a higher-order topological phase. Bulk-boundary correspondence states that an nD topological system holds (n-1)D boundary modes. However, boundary modes with lower dimensions such as (n-2)D also have been demonstrated recently \cite{Benalcazar61, serra2018observation}. This second-order topological phase has also been reported and experimentally verified in photonic crystals with 2D SSH lattice \cite{PhysRevLett.122.233902, PhysRevB.98.205147, PhysRevLett.122.233903}. In these systems, the existence of the corner modes requires an interface between topologically distinct crystals. However, the corner modes exist in the in-plane polarized 2D SSH model surrounded by vacuum, in other words, under an open boundary condition. To further investigate the corner mode, we considered in-plane polarized eigenmodes of an array of 20 by 20 unit cells (Fig. \ref{NbyN}(a)). It has four projected bulk states and three lower-dimensional states. Magnified views of three boundary states show doubly degenerate edge states (red and green boxes) and quadruply degenerate corner states (blue box). The normalized dipole moment distributions of edge and corner states are shown in Fig. \ref{NbyN}(b) and \ref{NbyN}(c). However, the same array supports only edge states (red and blue boxes in Fig. \ref{NbyN}(d)), which have normalized dipole momentum localized in $y$-parallel (Fig. \ref{NbyN}(e)) and $x$-parallel (Fig. \ref{NbyN}(f)) directions. The absence of the isolated corner mode of $z$ polarization originates from the $C_4$ symmetry that induces a degeneracy at $\Gamma$ and $M$. The degeneracy between the second and third bands gives rise to an overlap of projected bulk states and prevents the corner mode from being isolated from the bulk states.

\section{Discussion and conclusion}
Despite the topological features of the 2D SSH model, the sum of its Berry curvature over the first BZ is zero \cite{PhysRevLett.118.076803}. The topological phase of the 2D SSH model is instead characterized by the 2D Zak phase  \cite{PhysRevLett.118.076803}. The Zak phase \cite{PhysRevLett.62.2747}, associated with the shift of Wannier band, or bulk polarization, is quantized to $0$ for the trivial, and to $\pi$ for the nontrivial case. The extended 2D Zak phase $\theta$ can be calculated as
\begin{equation}
    \theta_j = -\frac{1}{2\pi}\int_{BZ}{d^2\vec{k} \text{Tr} [A_j (\vec{k})]}, \hspace{3mm} j = x, y.
    \label{Zak}
\end{equation}
Here, the ($m,n$) component of $A_j$ is given as $(A_j)_{mn}(\vec{k}) = i\langle{\vec{u}_{m\vec{k}}} |\partial k_j|\vec{u}_{n\vec{k}} \rangle$ where $|\vec{u}_{n\vec{k}} \rangle$ is the periodic part of the Bloch function of $n$-th band. To numerically obtain the 2D Zak phase, we use a Wilson loop \cite{Benalcazar61, PhysRevB.96.245115, PhysRevB.98.205147}. 
\begin{align}
    \theta_x = \frac{a}{2\pi}\int{dk_y v^n_x(k_y)} \notag \\
    \theta_y = \frac{a}{2\pi}\int{dk_x v^n_y(k_x)}
    \label{Wilson}
\end{align}
where $v^n_j$ is the $n$-th eigenvalue of $-i \log{\Pi_{m=0}^{M}[F_{j,\vec{k}+m\triangle k_j}]}$ for $M$ satisfying $(M+1)\triangle k_j = 2\pi/a$ and $j = x, y$; the ($m,n$) component of $F_{j,\vec{k}}$ is $(F_{j,\vec{k}})_{mn} = \int{\vec{u}^{*}_{m\vec{k}} (\vec{r}) \cdot \varepsilon(\vec{r}) \vec{u}_{n\vec{k}} (\vec{r})}$. As reported by previous publications \cite{PhysRevLett.118.076803, PhysRevB.98.205147}, the 2D Zak phase of $z$ polarization is $(\pi,\pi)$. The 2D Zak phase of in-plane polarization are also $(\pi,\pi)$, which reflects the existence of both $x$- and $y$-polarized edge modes, although they are hidden by the projected bulk states (Fig. \ref{edge}(d)-\ref{edge}(f)).

\begin{figure} \centering
\includegraphics [width=0.45\textwidth]{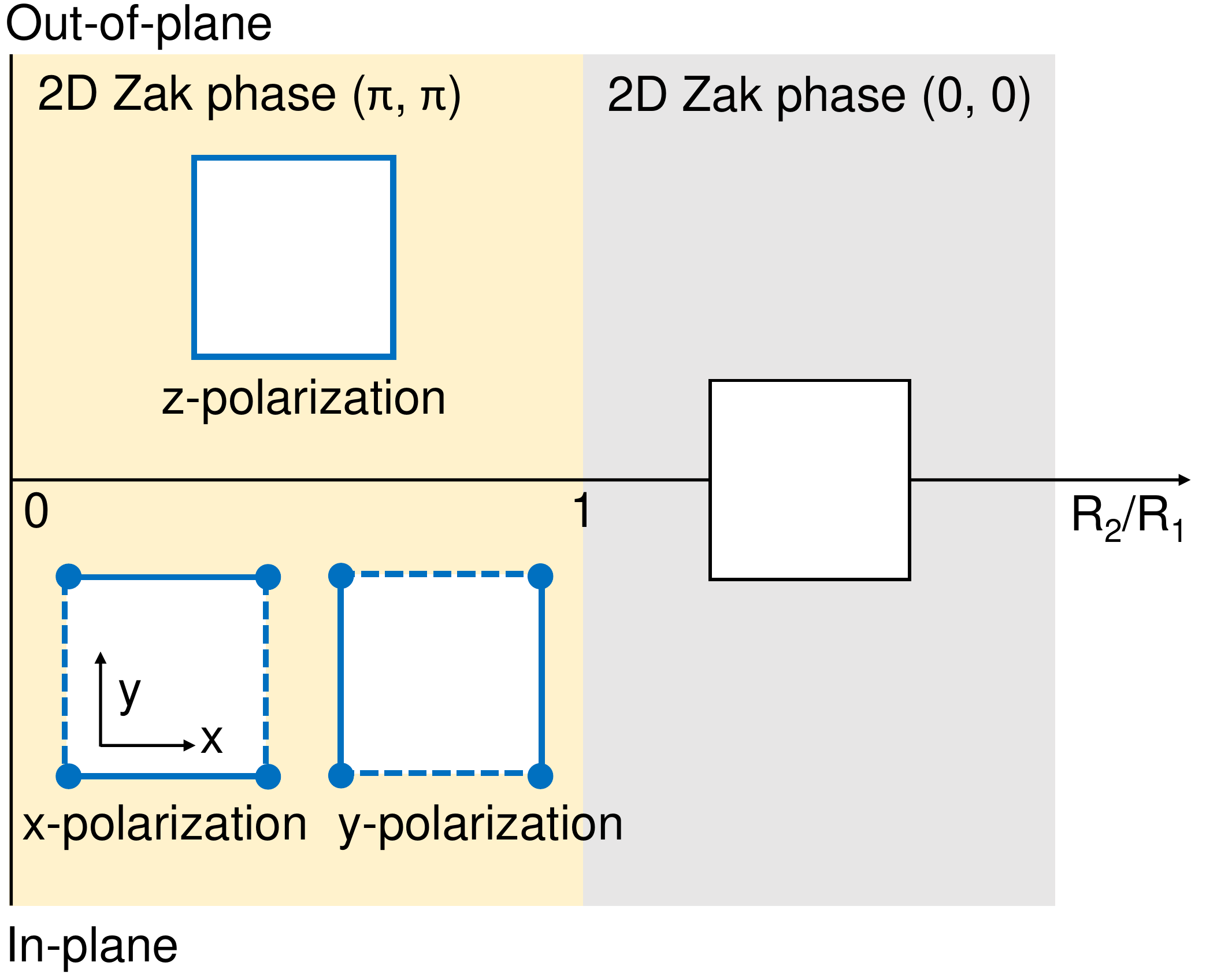}
\caption{Diagram of topological phases and edge/corner modes. $x$-axis represents ratio between intercellular and intracellular distance ($R_2/R_1$). $y$-axis distinguishes polarization of eigenmodes. Boxes and their boundaries correspond to arrays consisting of N by N unit cells and their edge modes under open boundary condition. Solid black line: no edge state, dashed blue line: edge state overlapped with bulk states, solid blue line: edge state. Blue dots at the corner represent the existence of corner modes.
}
\label{diagram}
\end{figure}

The topological phase and existence of edge and corner modes can be represented schematically (Fig. \ref{diagram}). Boxes correspond to arrays consisting of N by N unit cells under open boundary condition. Solid black (blue) lines indicate the absence (existence) of edge states while dashed blue lines represent edge states overlapped with the projected bulk states. Existence of the corner modes are marked by blue dots at the corner. When $R_1 > R_2$, the topological phase is nontrivial ($\theta = (\pi,\pi)$) whereas it is trivial ($\theta = (0,0)$) otherwise. The polarization direction does not affect the 2D Zak phase. However, the polarization determines the existence of isolated edge states and corner states. When the polarization is perpendicular to the 2D crystals, boundaries along both $x$- and $y$-axis support topological edge states that are isolated from the bulk states. Therefore, the edge states can be selectively excited while the bulk state remains insulating. For in-plane polarization, the same 2D SSH model supports edge states in all boundaries, but only longitudinal edge states are isolated from the bulk. The edge states propagating perpendicular to the polarization direction are hidden by the projected bulk states, thereby making it impossible to transport edge modes without affecting the bulk. Meanwhile, the corner modes exist only for in-plane polarization as shown in Fig. \ref{NbyN}. Therefore, polarization can be used as an additional degree of freedom to determine the existence of the corner modes.

In conclusion, we implemented a coupled dipole approximation to study polarization-dependent topological phase of an array of metallic nanoparticles in a 2D SSH lattice. By considering in-plane polarization, the existence of edge and corner modes were investigated. An in-plane polarized 2D SSH model possesses isolated longitudinal edge modes propagating parallel to the polarization direction. We also demonstrated that the nontrivial 2D SSH model under an open boundary condition supports a second-order topological phase by showing in-plane polarized corner modes, whereas previously reported corner modes have been formed between topologically distinct lattices. Using polarization as a new degree of freedom to control topological features, multifunctional photonic devices with robust wave control will become possible.
\begin{acknowledgments}
This work was financially supported by the National Research Foundation of Korea (NRF) (Grants No. NRF2019R1A2C3003129, No. CAMM-2019M3A6B3030637, No. NRF-2018M3D1A1058998, and No. NRF2015R1A5A1037668) funded by the Ministry of Science and ICT (MSIT) of the Korean government. M.K. acknowledge Global Ph.D. fellowship (NRF-2017H1A2A1043204) from NRF-MSIT of the Korean government

\end{acknowledgments}

\bibliographystyle{apsrev4-1}
\bibliography{\jobname}

\end{document}